\documentclass[sigconf]{acmart}
\usepackage{amsmath}
\usepackage{mathrsfs}
\usepackage{algorithm,algpseudocode}
\usepackage{color,xcolor}
\usepackage{algpseudocode}
\usepackage{amsmath}
\usepackage{balance}
\usepackage{graphics}
\usepackage{epsfig}
\usepackage{balance}
\usepackage{threeparttable}
\usepackage{enumitem}
\usepackage{bm}
\usepackage{verbatim}
\usepackage{soul}
\usepackage{multirow}
\usepackage{makecell}
\usepackage{xspace}
\usepackage{appendix}
\usepackage{booktabs}
\usepackage{multirow}
\usepackage[justification=justified,skip=7pt]{caption}

\newtheorem{myDef}{Definition}

\newcommand{\eg}{\emph{e.g.,}\xspace}
\newcommand{\ie}{\emph{i.e.,}\xspace}
\newcommand{\wrt}{\emph{w.r.t.}\xspace}
\newcommand{\model}{GRN}

\renewcommand{\arraystretch}{1.05}

\AtBeginDocument{%
  \providecommand\BibTeX{{%
    \normalfont B\kern-0.5em{\scshape i\kern-0.25em b}\kern-0.8em\TeX}}}

\setcopyright{acmcopyright}
\copyrightyear{2021}
\acmYear{2021}
\acmDOI{10.1145/1122445.1122456}

\acmPrice{15.00}
\acmISBN{978-1-4503-XXXX-X/18/06}

\begin{document}
\title{{\model}: Generative Rerank Network for Context-wise Recommendation}
\author{Yufei Feng, Binbin Hu, Yu Gong, Fei Sun, Qingwen Liu, Wenwu Ou}

\affiliation{%
\institution{Alibaba Group, Hangzhou, China} 
}
\email{
fyf649435349@gmail.com, bin.hbb@antfin.com, {gongyu.gy, ofey.sf, xiangsheng.lqw}@alibaba-inc.com, santong.oww@taobao.com
}

\begin{abstract}
Reranking is attracting incremental attention in the recommender systems, which rearranges the input ranking list into the final ranking list to better meet user demands. Most existing methods greedily rerank candidates through the rating scores from point-wise or list-wise models. 
Despite effectiveness, neglecting the mutual influence between each item and its contexts in the final ranking list often makes the greedy strategy based reranking methods sub-optimal.
In this work, we propose a new context-wise reranking framework named \textbf{G}enerative \textbf{R}erank \textbf{N}etwork ({\model} for short). Specifically, we first design the evaluator, which applies Bi-LSTM and self-attention mechanism to model the contextual information in the labeled final ranking list and predict the interaction probability of each item more precisely. Afterwards, we elaborate on the generator, equipped with GRU, attention mechanism and pointer network to select the item from the input ranking list step by step. Finally, we apply cross-entropy loss to train the evaluator and, subsequently, policy gradient to optimize the generator under the guidance of the evaluator.
Empirical results show that {\model} consistently and significantly outperforms state-of-the-art point-wise and list-wise methods. Moreover, {\model} has achieved a performance improvement of 5.2\% on PV and 6.1\% on IPV metric after the successful deployment in one popular recommendation scenario of Taobao application. 

\end{abstract}



\maketitle

\section{Introduction}

\begin{figure}
    \centering
    \includegraphics[scale=0.56]{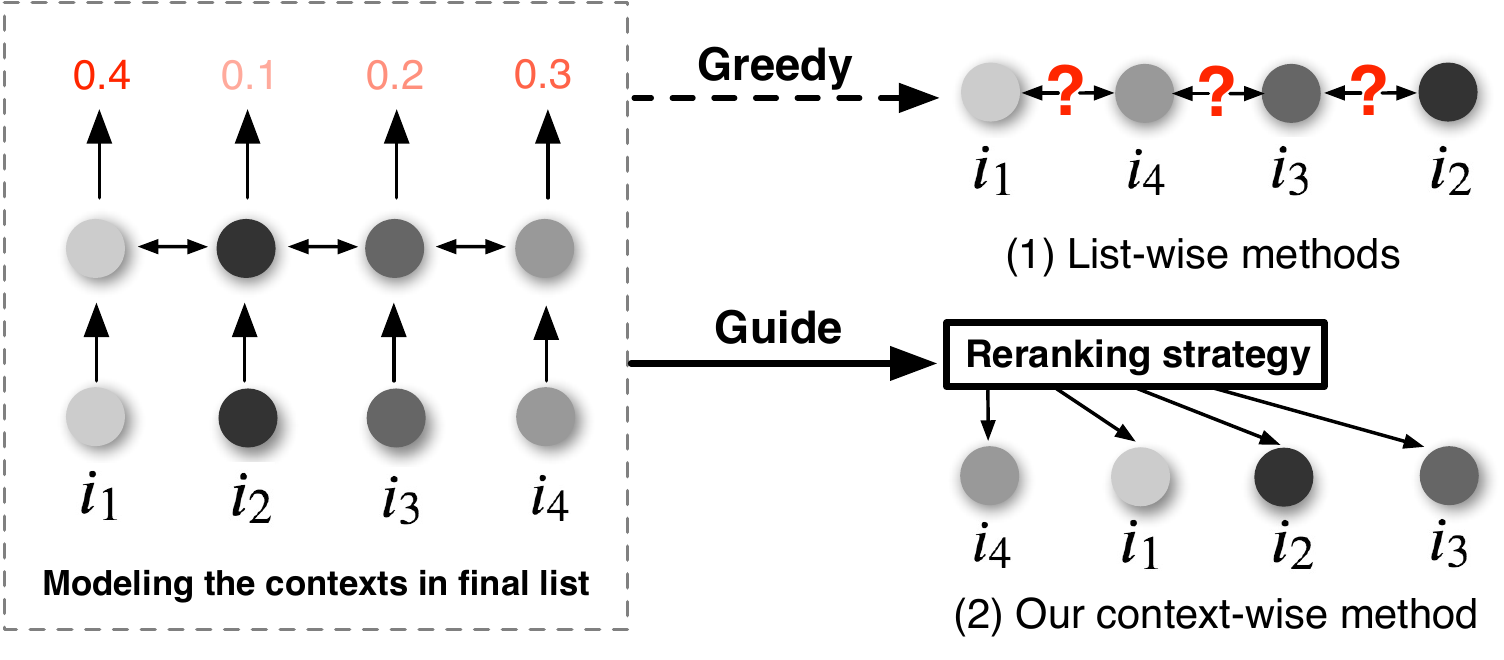}
    \caption{The difference of existing works and ours. Greedily reranking changes the original contexts and lead to inaccurate prediction in the new permutation. Our work attempts to learn a context-wise reranking strategy under the guidance of the well-trained context-wise model.}
    \label{fig:intro}
\end{figure}

Recommender systems (RS) have been widely deployed in various web-scale applications, including e-commerce~\citeN{Sarwar:ItemCF, Paul:YoutubeNet, Zhou:DIN, Feng:DSIN}, social media~\citeN{Paul:YoutubeNet, gomez2015netflix} and online news~\cite{das2007google, liu2010personalized, Li:SCENE}. Owing to the great impact on user
satisfaction as well as the revenue of the RS, recent attention is increasingly shirting towards the reranking stage, which is typically designed to rearrange the input ranking list generated by previous stages (i.e., matching and ranking).
With the rapid development of deep learning techniques, various reranking algorithms have been proposed to address the interior relevance in the input ranking list for improving recommendation performance. 


In particular, several recent efforts have attempted to follow the greedy strategy for reranking on items with refined rating scores, which mainly fall into two groups,
global point-wise models~\citeN{Paul:YoutubeNet, Cheng:WDL, Guo:DeepFM} and local list-wise models~\citeN{Ai:DLCM, Burges:lambdamart, Pei:PRM, Bello:Seq2slate, Zhuang:MIDNN}. 
Global point-wise models learn a ranking function with the labeled user-item pairs. In spite of effectiveness, they ignore different feature distributions within the input ranking list generated for each user. In fact, users may show different concerns about different input ranking lists, such as the price of the snacks and the brand of the electronics. 
To solve this, several local list-wise models have been proposed to refine the ranking scores by taking feature distributions of the input ranking list into account. Unfortunately, these methods commonly follow the greedy strategy for item rearrangements, which changes the contexts of each item from the initial list to the final list, resulting in imprecise estimation under the new item permutation.
As shown in Fig.~\ref{fig:intro}, the rating scores predicted by modeling the contexts of items $(i_1, i_2, i_3, i_4)$ in the final list could be $(0.4, 0.1, 0.2, 0.3)$. Specifically, $0.3$ represents the contextual interaction probability when placed before $i_2$ and after $i_1, i_4$. Once the candidates are rearranged according to the ratings scores to $(i_1, i_4, i_3, i_2)$, the contextual items of $i_3$ are modified, leading to inaccurate estimation in the new permutation and sub-optimal item arrangements. 

How to better leverage contexts in the reranking stage remains a great challenge in RS. 
Intuitively, instead of directly recommending a cheap item, placing a more expensive item ahead can stimulate the user's desire to buy the cheaper item. Besides, placing all the items of the user's most interest in the front may reduce the possibility of his/her continue browsing.
Such generalized contextual knowledge is of crucial importance in the item rearrangement. Unfortunately, such contexts are unable to determine and capture since the final ranking list is not given.
Inspired by continuous advances in actor-critic reinforcement learning~\citeN{schulman2015high, schulman2017proximal, mnih2016asynchronous}, one practical and effective solution is to learn a context-wise reranking strategy under the guidance of the well-trained context-wise model. Specifically, as illustrated in Fig.~\ref{fig:intro}, we transform the context-wise reranking objective into the following two tasks: 
(1) The local context-wise model is designed to predict the contextual interaction probability (\eg click-through rate and conversion rate) more precisely, by fully leveraging the contexts of each item from the labeled records between users and item lists; 
(2) The reranking strategy is meant to obtain the rearranged item list from the input ranking list. Moreover, the reranking strategy is converged to be context-wisely optimal under the guidance of the local context-wise model. In this way, by considering the contextual information, the reranking strategy can reach the context-wise item arrangements by distilling contextual knowledge from the local context-wise model, and then bring better experience for users.

In this work, with the above discussions in mind, we propose a novel context-wise framework named \textbf{G}enerative \textbf{R}erank \textbf{N}etwork ({\model} for short), which is composed of three parts: 
(1) Evaluator. To better refine the contextual interaction probability of each item in the final ranking list, we employ Bi-LSTM and self-attention mechanism to capture sequential information alongside the list and mutual influence between items, respectively; 
(2) Generator. With the aim of learning the reranking strategy to generate the final ranking list, we apply pointer network combined with GRU and attention mechanism to select appropriate item from the input ranking list at each step;
(3) Training procedure. We adopt cross-entropy loss to train evaluator and, subsequently, policy gradient with technically designed advantage reward to train the generator. 

In sum, we make the following contributions:
\begin{itemize}
    \item We highlight the necessity of modeling the contexts in the final item list to each item for more precise prediction and better item rearrangements. We propose a practical and effective solution is to learn a context-wise reranking strategy under the guidance of the well-trained context-wise model. 
    \item We propose a novel context-wise reranking framework named {\model}, which simultaneous employs the  well-designed evaluator and generator in an cooperative manner,  with the aims of comprehensively capturing evolving intent and mutual influence implied in input and final ranking list.
    \item We perform a series of experiments on a public dataset from Alimama and a proprietary dataset from Taobao application. Online and offline experimental results demonstrate that {\model} consistently and significantly outperforms various state-of-the-art methods.
\end{itemize}

The remainder of this paper is organized as follows. Section 2 discusses related work for reranking, including point-wise, pair-wise and list-wise models. Section 3 presents some preliminary knowledge as well as the task description. Then, we propose the generative rerank network in Section 4. Experiments and detailed analysis are reported in Section 5. Finally, we conclude the paper in Section 6.

\section{Related Work}
Reranking usually serves as the last stage after matching and ranking stages in one typical recommender system. In this section, we review the most related studies in the reranking stage of recommender systems, where an input ranking list provided by previous stages is rearranged into the final ranking list and expected to better meet user demands. Most reranking methods can be broadly classified into three categories: point-wise, pair-wise and list-wise methods. 
\begin{itemize}
    \item \textbf{Point-wise} methods~\citeN{Paul:YoutubeNet, Cheng:WDL, Guo:DeepFM} regard the recommendation task as a binary classification problem and globally learn a scoring function for a given user-item pair with manually feature engineering. Though with continuous achievements, these methods neglect considering the comparison and mutual influence between items in the input ranking list.
    
    \item \textbf{Pair-wise} models work by considering the semantic distance of an item pair every time. Following this line, a surge of works are proposed to techinically designed pair-wise loss functions to compare any item pair in the input ranking list with well-designed architecture, including  RankSVM~\cite{Joachims:RankSVM} based on SVM, GBRank~\cite{Zheng:GBRank} based on GBDT as well as  RankNet~\cite{Burges:RankNet} and DSF~\cite{ai2019learning}
    based on emerging deep neural networks. 
    However, pair-wise methods neglect the local information in the list and increase the model complexity.
    
    \item \textbf{List-wise} models are proposed to capture the interior correlations among items in the list in different ways.
    LambdaMart~\cite{Burges:lambdamart} is a well-known tree-based method with the list-wise loss function.
    MIDNN~\cite{Zhuang:MIDNN} works by handmade global list-wise features, while it requires much domain knowledge and decreases its generalization performance. DLCM~\cite{Ai:DLCM} and PRM~\cite{Pei:PRM} apply GRU and transformer to encode the list-wise information of the input ranking list for better prediction, respectively. 
    Though effective, this type of list-wise models do not escape the paradigm of greedy ranking based on predicted scores, where the adjustment of the final ranking list greatly changes the contextual information of each item. 
\end{itemize}
Most closest to our work are MIRNN~\cite{Zhuang:MIDNN} and Seq2slate~\cite{Bello:Seq2slate}, which directly learns a step-greedy reranking strategy to generate the final ranking list through RNN and pointer network, respectively. 
We argue that only considering the preceding information is insufficient and incomplete for the optimal item rearrangements since the interaction probability of each item is heavily affected by both preceding and subsequent ones. In this work, we upgrade the step-greedy strategy based reranking methods to the context-wise reranking strategy. 

There are also some works~\cite{Chen:DPP, Wilhelm:DPPInYoutube, Gelada:DeepMDP, Gogna:BalancingAccAndDiversity} focusing on making the trade-off between relevance and diversity in the reranking stage. Different from these works, {\model} is an end-to-end context-wise reranking framework, which may automatically generate diverse or alike recommendaion results for the only sake of effectiveness. Besides, group-wise methods~\cite{ai2019learning} determine the priority of items by multiple documents in the list. Though effective, the computation complexity of group-wise methods is at least $O(N^2)$, which may not be appropriate for industrial recommender systems.
\section{Preliminary}

In general, a mature recommender system (\eg e-commerce and news) is composed of three stages chronologically: matching, ranking and reranking. In this paper, we focus on the final reranking stage, whose input is the ranking list produced by the previous two stage (\ie matching and ranking). The goal of the reranking is to elaborately select candidates from the input ranking list and rearrange them into the final ranking list, followed by the exhibition for users.
Mathematically, with user set $\mathcal{U}$ and item set $\mathcal{I}$, we denote list interaction records as $\mathcal{R} = \{(u, \mathcal{C}, \mathcal{V}, \mathcal{Y} | u \in \mathcal{U}, \mathcal{V} \subset \mathcal{C} \subset \mathcal{I}\}$. Here, $\mathcal{C}$ and $\mathcal{V}$ represent the recorded input ranking list with $m$ items for reranking stage and the recorded final ranking list with $n$ items exhibited to user $u$, respectively. Intuitively, $n \leq m$. $y^t_{uv} \in \mathcal{Y}$ is the implicit feedback of user $u$ \wrt $t$-th item $v_t \in \mathcal{V}$, which can be defined as follows:
$y^t_{uv} = 1$ when interaction (\eg click) is observed, and $y^t_{uv} = 0$ otherwise. 
In the real-world industrial recommender systems, each user $u$ is associated with a user profile $x_u$ consisting of sparse features $x^u_s$ (\eg user id and gender) and dense features $x^u_d$ (\eg age), while each item $i$ is also associated with a item profile $x_i$ consisting of sparse features $x^s_i$ (\eg item id and brand) and dense features $x_i^d$ (\eg price) .


Given the above definitions, we now formulate the reranking task to be addressed in this paper:
\begin{myDef}
\textbf{Task Description}. Given a certain user $u$, involving his/her input ranking list $\mathcal{C}$, the goal is to learn a reranking strategy $\pi: \mathcal{C} \stackrel{\pi}{\longrightarrow} \mathcal{O}$, which aims to select and rearrange items from $\mathcal{C}$, and subsequently recommend a final ranking list $\mathcal{O}$ that are of the most interest to $u$.
\end{myDef}

Many efforts have been made for reranking task, while most of these works ignore the contextual information in the final ranking list (i.e., $\mathcal{V}$), resulting in sub-optimal performance. In practice, such contextual knowledge is of crucial importance, since users are commonly sensitive to the permutation of items during browsing.


\section{The Proposed Framework}

\begin{table}
\renewcommand{\arraystretch}{1.2}
\centering
  \caption{Key notations.}
  \label{table:notations}
  \begin{tabular}{ll}
    \toprule
    Notations & Description\\
    \midrule
    $\mathcal{U}$, $\mathcal{I}$ & \makecell[{}{p{5.8cm}}]{the set of users and items, respectively}\\
    \midrule
    $\mathcal{C}$, $\mathcal{V}$ & \makecell[{}{p{5.8cm}}]{the recorded input and final ranking list, respectively}\\
    \midrule
    $\mathcal{Y}$ & \makecell[{}{p{5.8cm}}]{the implicit feedback towards the ranking list}\\
    \midrule
    $m, n$ & \makecell[{}{p{5.8cm}}]{the number of items in the recorded input and final ranking list, respectively}\\
    \midrule
    $\pi$ & \makecell[{}{p{5.8cm}}]{the learned reranking strategy}\\
    \midrule
     $\mathcal{O}$ & \makecell[{}{p{5.8cm}}]{the generated final ranking list}\\
    \midrule
    $x_s^u , x_d^u$ & \makecell[{}{p{5.8cm}}]{the sparse and dense feature set for users, respectively} \\
    \midrule
    $x_s^i, x_s^i$ & \makecell[{}{p{5.8cm}}]{the sparse and dense feature set for items, respectively} \\
    \midrule
    
     $\mathbf{\Theta}^G$, $\mathbf{\Theta}^E$ & \makecell[{}{p{5.8cm}}]{parameters of generator and evaluator, respectively}\\
    \bottomrule
  \end{tabular}
\end{table}

In this section, we introduce our proposed framework {\model}, which aims to achieve context-wise item arrangements in the reranking stage of recommender systems. Overall, {\model} is composed of three parts: (1) To predict the contextual probability of being interacted in the labeled final ranking list $\mathcal{V}$, we propose the local context-wise model named evaluator, where Bi-LSTM and self-attention mechanism are applied to capture the sequential information and mutual influence between items; (2) To generate the final ranking list $\mathcal{O}$ from the input ranking list $\mathcal{C}$, we elaborate on the model design of generator. Specifically, we equip the generator with pointer network, GRU and attention mechanism, with the hope of the abilities to select the most suitable item from $\mathcal{C}$ and capture adjacent information; (3) To converge the generator to a preeminent reranking strategy under the guidance of evaluator, we pay special attention to designing the training procedure. First, we adopt cross-entropy loss to train the evaluator with labeled list records. Afterwards, policy gradient cooperated with proposed advantage reward are applied to train the generator. The key notations we will use throughout the article are summarized in Table \ref{table:notations}.

First of all, we begin with the representations of users and items, which are basic inputs of our proposed model. Following previous works~\cite{Paul:YoutubeNet, Zhou:DIN}, we parameterize\footnote{Note that the parameters are not shared in the evaluator and generator.} the available profiles into vector representations for users and items. Given a user $u$, associated with sparse features $x_u^s$ and dense features $x_u^d$,  we embed each sparse feature value into $d$-dimensional space, while handle dense features standardization or batch normalization to ensure normal distribution. Subsequently, each user $u$ can be represented as $\mathbf{x}_u \in \mathbb{R}^{|x^s_u| \times d + |x^d_u|}$, where $|x^s_u|$ and $|x^d_u|$ denote the size of sparse and dense feature space of user $u$, respectively. Similarly, we represent each item $i$ as $\mathbf{x}_i \in \mathbb{R}^{|x^s_i| \times d + |x^d_i|}$. Naturally, we represent the input ranking list $\mathcal{C}$ as $\mathbf{C} = [\mathbf{x}_c^1, ..., \mathbf{x}_c^m]$ and the final ranking list  $\mathcal{V}$ as $\mathbf{V} = [\mathbf{x}_v^1, ..., \mathbf{x}_v^n]$, where $m$ and $n$ is the number of items in the input ranking list and final ranking list, respectively.

In the ensuing sections, we shall zoom into the proposed evaluator and generator for {\model}.

\begin{figure}
    \centering
    \includegraphics[scale=0.45]{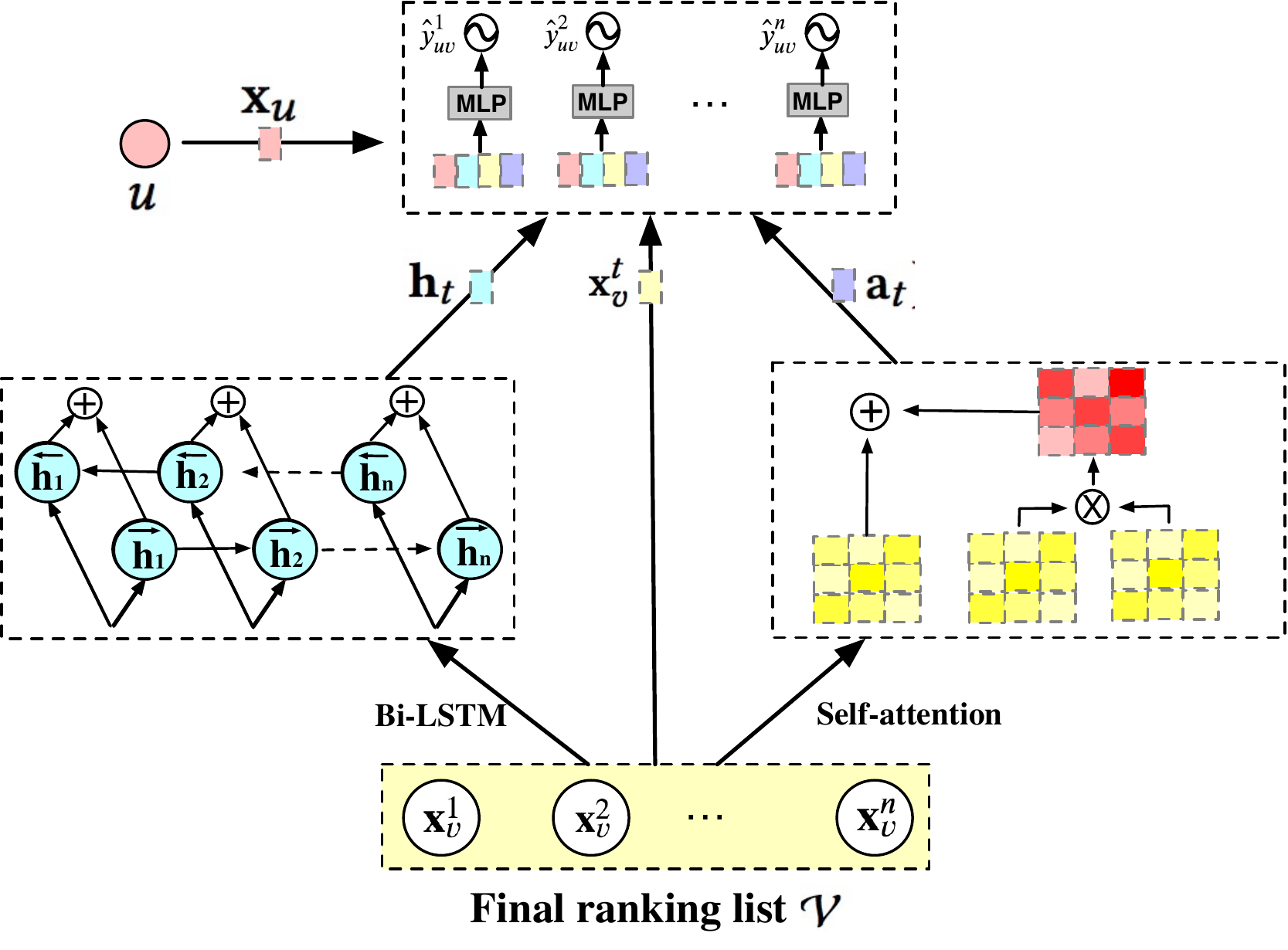}
    \caption{The overall architecture of evaluator in {\model}.}
    \label{fig:Evaluator}
\end{figure}

\subsection{Evaluator}
As motivated, in the context-wise view, the underlying reasons driving a user to interact with a item in a list may depend on following two aspects:
(1) the two-way evolution of user's intent during browsing;
(2) the mutual influence between items in the item list.
Thus, by taking above factors into consideration, our evaluator $E(\mathbf{x}_v^t|u, \mathcal{V}; \mathbf{\Theta}^{E})$, parameterized by $ \mathbf{\Theta}^{E}$, estimates the contextual interaction probability between user $u$ and the $t$-th item $\mathbf{x}_v^t$ in the final ranking list $\mathcal{V}$. Specifically, our evaluator aims to capture following two aspects of information implicated in sequences.
\begin{itemize}
    \item \textbf{Intent evolution.} Naturally, user intent keeps bi-directionally evolving when browsing items sequentially, and thus, it is imperative to capture such dynamics of intent for modeling user preference. Here, we adopt the commonly used Bi-LSTM~ \cite{Hochreiter:LSTM} to deal with sequences for long-term and short-term preference.  Mathematically, the forward output state for the $t$-th item $x_v^t$ can be calculated as follows:
\begin{equation} \label{eq:eva-bi-lstm}
\begin{aligned}
\textbf{i}_t &= \sigma(\textbf{W}_{xi}\mathbf{x}_v^t + \textbf{W}_{hi}\textbf{h}_{t-1} + \textbf{W}_{ci}\textbf{c}_{t-1} + \textbf{b}_i)\\
\textbf{f}_t &= \sigma(\textbf{W}_{xf}\mathbf{x}_v^t + \textbf{W}_{hf}\textbf{h}_{t-1} + \textbf{W}_{cf}\textbf{c}_{t-1} + \textbf{b}_f)\\
\textbf{c}_t &= \textbf{f}_{t}\mathbf{x}_v^t + \textbf{i}_{t}\tanh(\textbf{W}_{xc}\mathbf{x}_v^t + \textbf{W}_{hc}\textbf{h}_{t-1} + \textbf{b}_{c})\\
\textbf{o}_t &= \sigma(\textbf{W}_{xo}\mathbf{x}_v^t + \textbf{W}_{ho}\textbf{h}_{t-1} + \textbf{W}_{co}\textbf{c}_{t} + \textbf{b}_o)\\
\overrightarrow{\textbf{h}_{t}} &= \textbf{o}_{t}\tanh(\textbf{c}_t)\\
\end{aligned}
\end{equation}
where $\sigma(\cdot)$ is the logistic function, and $\textbf{i}$, $\textbf{f}$, $\textbf{o}$ and $\textbf{c}$ are the input gate, forget gate, output gate and cell vectors which have the same size as $\mathbf{x}_v^t$. Shapes of weight matrices are indicated with the subscripts. Similarly, we can get the backward output state $\overleftarrow{\textbf{h}_{t}}$. Then we concatenate the two output states $\textbf{h}_{t} = \overrightarrow{\textbf{h}_{t}} \oplus \overleftarrow{\textbf{h}_{t}}$ and denote $\textbf{h}_{t}$ as the sequential representation of $x_v^t$.   
    
    \item \textbf{Mutual influence.} That is, any two items can affect each other, although they are far away from each other in the final ranking list. To achieve this goal, we apply the recently emerging self-attention mechanism~\cite{Vaswani:Transformer} to directly model the mutual influences for any two items regardless the distances between them. Formally, we implement it as follows:
\begin{equation} \label{eq:eva-self-att}
\begin{split}
\mathbf{A} = \mbox{softmax}\left(\frac{\mathbf{V}\mathbf{V}^{T}}{\sqrt{d_k}}\right)\mathbf{V}
\end{split}
\end{equation}
where $d_k$ is the dimension of each item representation in $\mathcal{V}$. Obviously, we can easily extend it to multi-head attention for the stable training process. We denote $\mathbf{a}_{t}$, the $t$-th representation in $\mathbf{A}$, as the mutual representation of $x_v^t$.
\end{itemize} 

Due to the powerful ability in modeling complex interaction in the CTR prediction field~\citeN{Zhou:DIN,Feng:DSIN,Pi:MIMN,zhou2019deep}, we integrate the multi-layer perceptron (MLP) into the evaluator for better feature interaction. Hence, we formalize our evaluator as follows:
\begin{equation} \label{eq:eva-mlp}
    \begin{split}
    E(\mathbf{x}_v^t|u, \mathcal{V}; \mathbf{\Theta}^{E}) = \sigma\bigl(f(f(f(\mathbf{x}_u \oplus \mathbf{x}_v^t \oplus \textbf{h}_{t} \oplus \mathbf{a}_{t})))\bigr)
    \end{split}
\end{equation}
where $f(\textbf{x}) = ReLU(\textbf{W}\textbf{x} + \textbf{b})$, $\sigma(\cdot)$ is the logistic function. The parameter set of the evaluator is thus $\mathbf{\Theta}^{E}= \{\mathbf{W}_*, \mathbf{b}_*\}$, \ie the union of parameters for Bi-LSTM and MLP.

Clearly, our evaluator can be optimized via binary cross-entropy loss function, which is defined as follows:
\begin{equation} \label{eq:eva-loss}
    \begin{split}
    \mathcal{L}^{E} = -\frac{1}{N}\!\sum_{(u, \mathcal{V}) \in \mathcal{R}}\sum_{x_v^t \in \mathcal{V}} \left(y_{uv}^t\,\log \hat{y}^t_{uv} + (1 {-} y_{uv}^t)\,\log(1 {-} \hat{y}_{uv}^t)\right)
    \end{split}
\end{equation}
where $\mathbb{D}$ is the training dataset. For convenience, we refer $\hat{y}_{uv}^t$ as $E(\mathbf{x}_v^t|u, \mathcal{V}; \mathbf{\Theta}^{E})$, \ie $\hat{y}_{uv}^t =  E(\mathbf{x}_v^t|u, \mathcal{V}; \mathbf{\Theta}^{E})$, and $y_{uv}^t \in \{0, 1\}$ is the ground truth. We can optimize the parameters $\mathbf{\Theta}^E$ through minimizing $\mathcal{L}^E$.

\begin{figure}
    \centering
    \includegraphics[scale=0.5]{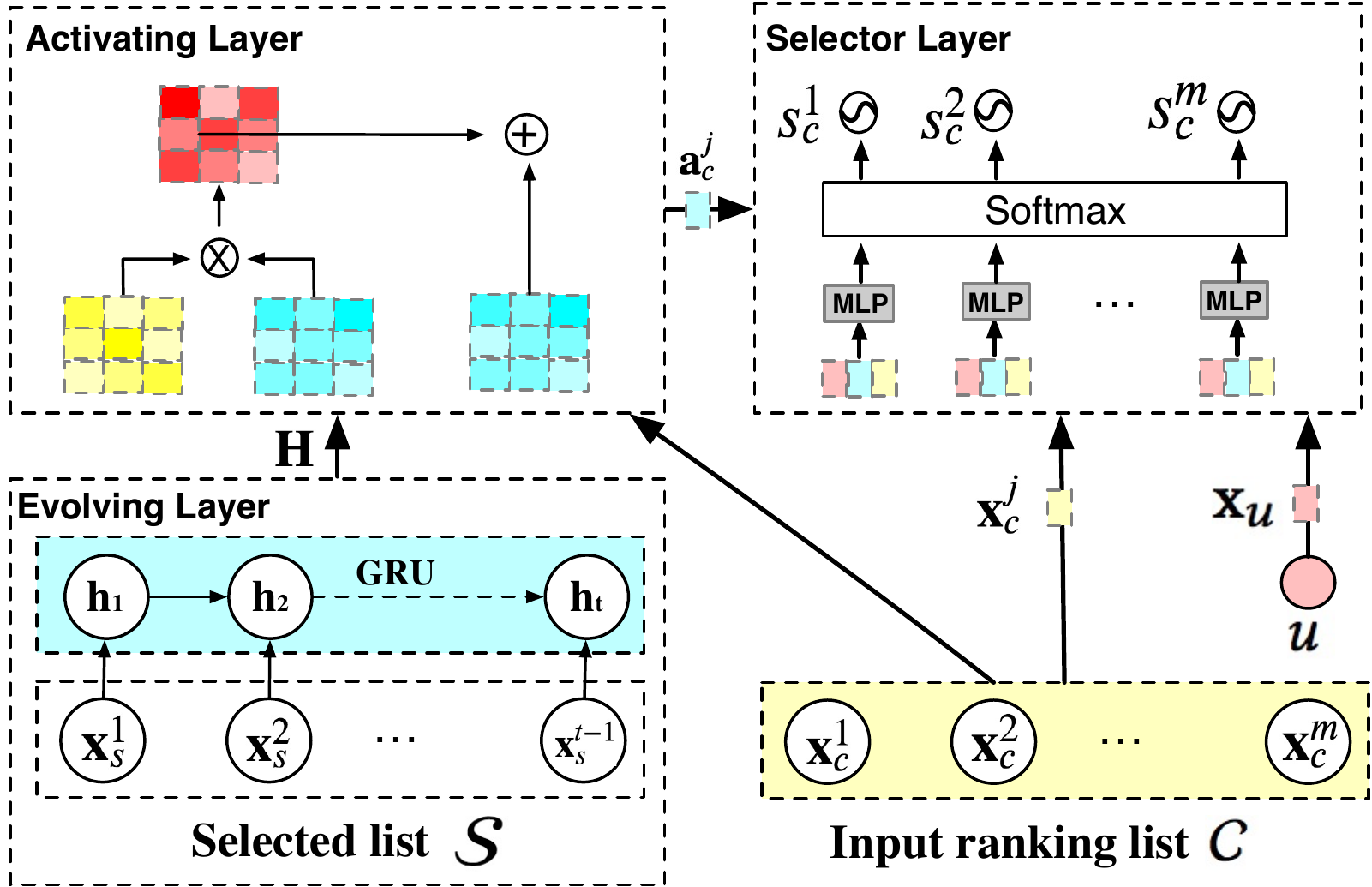}
    \caption{The overall architecture of generator in {\model}.}
    \label{fig:Generator}
\end{figure}

\subsection{Generator}
The goal of our generator $G(u, \mathcal{C}; \mathbf{\Theta}^G)$, parameterized by $ \mathbf{\Theta}^G$, is to learn a reranking strategy for item selection from input ranking list $\mathcal{C}$, that are of the most interest to user $u$. In the following sections, we focus on how to select the appropriate item at each step by considering the user's dynamic state during browsing, besides how to generate the final ranking list and train the generator.

\textbf{Evolving Layer.} As motivated, user's intent or interest will change over time, which derive us to learn dynamic representation for the target user to adapt for the evolution of the browsing history. For this purpose, we aim to distill the sequential information into the state representation at each step for the target user $u$, which achieved by the \textit{time-efficient} GRU module. Formally, at the $t$-th step, let $\mathcal{S} = [\mathbf{x}_s^0, \mathbf{x}_s^1, ..., \mathbf{x}_s^{t-1}]$ be the $t-1$ items selected from $\mathcal{C}$~\footnote{Specially, for the first step, $\mathbf{x}_l^0$ is a trainable vector which has the same embedding size with $\mathbf{x}_l^t$}, the output of GRU module can be calculated as follows,

\begin{equation}\label{eq:gen-gru}
    \begin{split}
    \textbf{z}_t &= \sigma(\textbf{W}_z \mathbf{x}_l^{t-1} + \textbf{U}_z \textbf{h}_{t-1}) \\
    \textbf{r}_t &= \sigma(\textbf{W}_t \mathbf{x}_l^{t-1} + \textbf{U}_t \textbf{h}_{t-1}) \\
    \widetilde{\textbf{h}}_t &= \text{tanh}(\textbf{W}\mathbf{x}_l^{t-1} + \textbf{U}(\textbf{r}_t\textbf{h}_{t-1})) \\
    \textbf{h}_t &= (1 - \textbf{z}_t)\textbf{h}_{t-1} + \textbf{z}_t\widetilde{\textbf{h}}_t 
    \end{split}
\end{equation}
where $\textbf{W}_z, \textbf{U}_z, \textbf{W}_t, \textbf{U}_t, \textbf{W}$ and $\textbf{U}$ are trainable variables. We denote $\mathbf{H} = [\mathbf{h}_1, ..., \mathbf{h}_t]$ as the sequential encoding of the selected list $\mathcal{S}$.

\textbf{Activating Layer.} Typically, the selected items have different influence towards the remaining items in the input ranking list. It is a common phenomenon that a user may get tired of the clothing category after browsing abundant different clothes. Hence, in this case, recommending other categories matching his/her interest could be a better choice. Inspired by vanilla attention mechanism~\cite{Dzmitry:Attention} , we learn the attention weights over remaining items in the input ranking list conditioned on the sequential representation of selected items. Given the representation of $j$-th remaining item $\mathbf{x}_c^j$ and the sequential representations of the selected list $\mathbf{H}$, we adopt weighted sum to generate the new representation for the $j$-th in the input ranking list with the corresponding attention weights as follows:
\begin{equation}\label{eq:gen-att}
    \begin{split}
    a^i_h &= \frac{\exp(\mathbf{h}_{i}\mathbf{W}\mathbf{x}_c^j)}{\sum_{i=1}^{t} \exp(\mathbf{h}_{i}\mathbf{W}\mathbf{x}_c^j)} \\
    \mathbf{a}^j_c &= \sum_{i=1}^{t} a^i_h\mathbf{h}_{i}\\
    \end{split}
\end{equation}

\textbf{Selector Layer.} After obtaining the representations for items (\ie $\mathbf{a}^j_c$) in input ranking list, we now aim to select the most suitable item, which best matches the potential interest for the target user at the $t$-th step. Here, we choose pointer network~\cite{Vinyals:Ptr} to achieve this goal.  Formally, for the $j$-th item $\mathbf{x}_c^j$ in the input ranking list $\mathbf{C}$, we first apply MLP for better feature interaction at the $t$-th step:
\begin{equation}\label{eq:gen-mlp}
    \begin{split}
    \widetilde{s}_c^j = f(f(f(\mathbf{x}_u \oplus \mathbf{x}_c^j \oplus \mathbf{a}^j_c)))
    \end{split}
\end{equation}
Afterwards, we apply the softmax function to normalize the vector $\widetilde{s}_c^j$ as follows:
\begin{equation}\label{generator softmax}
    \begin{split}
    s_c^j = \frac{\exp(\widetilde{s}_{c}^{j})}{\sum_{j}^{m} \exp(\widetilde{s}_{c}^{j})} \\
    \end{split}
\end{equation}
where $m$ is the number of items in $\mathbf{C}$. Then, at $t$-th step, our generator will select the item with the highest selection probability excluding the selected items, which can be formalized as follows:
\begin{equation}\label{eq:gen-select}
    G^t(u, \mathcal{C}; \mathbf{\Theta}^G) = x^{\mathop{\arg\max}_j{s^j_c}}_c \ \ \ \ s.t.\  x^{\mathop{\arg\max}_j{s^j_c}}_c \notin \mathcal{S}
\end{equation}
The parameter set of the generator is $\mathbf{\Theta}^G = \{\mathbf{U}_*, \mathbf{W}_*, \mathbf{b}_*\}$, \ie the union of parameters for GRU module and point network.

\textbf{List Generation.} Our generator is recurrently repeated until the generated final ranking list reaches the pre-defined length (\ie $n$). After repeating calculating $G^t(u, \mathcal{C}; \mathbf{\Theta}^G)$ for $n$ times, we get the reranking strategy $\pi = [G^1(u, \mathcal{C}; \mathbf{\Theta}^G), ..., G^n(u, \mathcal{C}; \mathbf{\Theta}^G)]$ and the generated final ranking list $\mathcal{O} = [x_o^1, ..., x_o^n]$, correspondingly.

\textbf{Advantage Reward.} As illustrated earlier, we utilize the evaluator to guide the optimization of the generator for the context-wise reranking strategy through policy gradient. As mentioned above, at each step, our generator will select the most suitable item from $\mathcal{C}$ in the light of the probability distribution in Eq.~\ref{generator softmax}. Besides, we propose the \textit{advantage reward} to estimate the actual reward of each item in the final ranking list more comprehensively. Specifically, the advantage reward consists of two parts: 
\begin{itemize}
    \item \textbf{Self reward}: As motivated, the interaction probability of itself in the final ranking list heavily affected by its context. Here we use the predicted contextual score by the evaluator to denote the its own reward in the list, which is calculated as follows:
    \begin{equation} \label{eq:advantage-reward-1}
    \begin{split}
    r^{self}(x_o^t|u, \mathcal{O}) &= \mbox{E}(x_o^t|u, \mathcal{O}; \mathbf{\Theta}^E) \\
    \end{split}
    \end{equation}
    \item \textbf{Differential reward}: Comprehensively, besides the self reward, each item brings differences in the interaction probability of other items in the list. For example, though selecting an unattractive item is generally not a good idea, it is appreciated if it can increase the interaction probability of other items. Hence, we propose the differential reward, with the assistance of the generator and calculated as follows:
    \begin{equation}  \label{eq:advantage-reward-2}
    \begin{split}
    r^{diff}(x_o^t|u, \mathcal{O}) = &\sum_{x_o^i \in \mathcal{O}^{-}}\mbox{E}(x_o^i|u, \mathcal{O}; \mathbf{\Theta}^E) - \\ &\sum_{x_o^i \in \mathcal{O}^{-}}\mbox{E}(x_o^i|u, \mathcal{O}^{-}; \mathbf{\Theta}^E) \\
    \end{split}
\end{equation}
where $\mathcal{O}^{-}$ is the item list which removes $x_o^t$ from $\mathcal{O}$.
\end{itemize}
 By combining both above rewards, we define the advantage reward of $t$-th item $x_o^t$ in $\mathcal{O}$ as follows:
\begin{equation}  \label{eq:advantage-reward-3}
    \begin{split}
    &r(x_o^t|u, \mathcal{O}) = r^{self}(x_o^t|u, \mathcal{O}) + r^{diff}(x_o^t|u, \mathcal{O})\\
    &= (\mbox{E}(x_o^t|u, \mathcal{O}; \mathbf{\Theta}^E) + \sum_{x_o^i \in \mathcal{O}^{-}}\mbox{E}(x_o^i|u, \mathcal{O}; \mathbf{\Theta}^E)) - \sum_{x_o^i \in \mathcal{O}^{-}}\mbox{E}(x_o^i|u, \mathcal{O}^{-}; \mathbf{\Theta}^E) \\
    &= \sum_{x_o^i \in \mathcal{O}}\mbox{E}(x_o^i|u, \mathcal{O}; \mathbf{\Theta}^E) - \sum_{x_o^i \in \mathcal{O}^{-}}\mbox{E}(x_o^i|u, \mathcal{O}^{-}; \mathbf{\Theta}^E)
    \end{split}
\end{equation}
Finally, the loss function of the generator is defined as follows:
\begin{equation} \label{eq:generator loss function}
    \begin{split}
    \mathcal{L}^{G} = -\frac{1}{N}\sum_{(u, \mathcal{C}) \in \mathcal{R}}\sum_{x_o^t \in \mathcal{O}} r(x_o^t|u, \mathcal{O})\log s_c^j.
    \end{split}
\end{equation}
Here $s_c^j$ represents the sampling probability of $x_o^t$ from $\mathcal{C}$ in Eq.~\ref{generator softmax}. We can optimize the parameters $\mathbf{\Theta}^G$ through minimizing $\mathcal{L}^G$.

\begin{figure}
    \centering
    \includegraphics[scale=0.85]{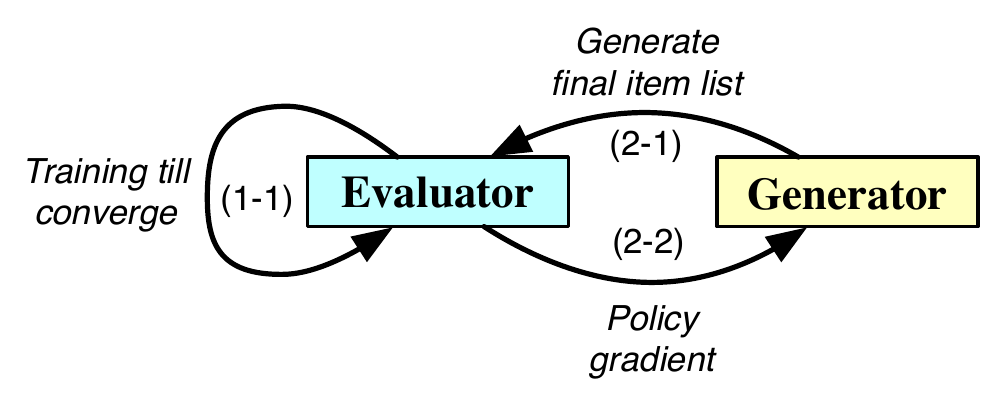}
    \caption{The illustration of the training procedure for {\model}.}
    \label{fig:Training procedure}
\end{figure}

\begin{algorithm}[t]
\caption{Training procedure for {\model}} 
\label{alg:optimization} 
\begin{algorithmic}[1] 
\Require 
List interaction records as $\mathcal{R} = \{(u, \mathcal{V}, \mathcal{C}, \mathcal{Y}) \}$; Evaluator $E(\mathbf{x}_v^t|u, \mathcal{V}; \mathbf{\Theta}^{E})$; Generator $G(u, \mathcal{C}; \mathbf{\Theta}^G)$
\Ensure Converged parameters $\mathbf{\Theta}^{E}$ and $\mathbf{\Theta}^G$;
\State $//$ \textbf{Evaluator training}
\While{$\mathbf{\Theta}^{E}$ not converge}
    \State Calculate prediction scores of evaluator by Eq.~\ref{eq:eva-bi-lstm} $\sim$ \ref{eq:eva-mlp};
    \State Optimizing $\mathbf{\Theta}^{E}$ with loss in Eq. \ref{eq:eva-loss}
\EndWhile 
\State $//$ \textbf{Generator training}
\While{$\mathbf{\Theta}^{G}$ not converge}
    \State New final ranking list $\mathcal{O}$;
    \For{$t = 1, 2, ..., n$}:
        \State Generate the $t$-th item $x_o^t$ from $\mathcal{C}$ by Eq.~\ref{eq:gen-gru} $\sim$ \ref{eq:gen-select};
        \State $\mathcal{O} \leftarrow \mathcal{O} \cup x_o^t$;
    \EndFor
    \State Calculate the advantage reward of each item in $\mathcal{O}$ by Eq.~\ref{eq:advantage-reward-1} $\sim$ \ref{eq:advantage-reward-3};
    \State Optimize $\mathbf{\Theta}^{G}$ with loss in Eq. \ref{eq:generator loss function}
\EndWhile 
\end{algorithmic}
\end{algorithm}

\subsection{Training Procedure}
we adopt the two-stage optimization strategy to train {\model}, which is illustrated in Fig.~\ref{fig:Training procedure}. We first utilize the list interaction records to optimize $\mathbf{\Theta}^E$ and thus improve the evaluator until converge ((1-1) in Fig.~\ref{fig:Training procedure}). Next, we fix $\mathbf{\Theta}^E$, and optimize $\mathbf{\Theta}^G$ in order to produce the final ranking list and better meet user's demands. Specifically, the generator will produce the final ranking list for the evaluator((2-1) in Fig.~\ref{fig:Training procedure}), and then update the paratemer $\mathbf{\Theta}^G$ via policy gradient with advantage reward based on the evaluator((2-2) in Fig.~\ref{fig:Training procedure}). The above process is repeated for more iterations until our generator converges. Overall, the training procedure for {\model} is outlined in Algorithm 1.

\section{Experiments}
In this section, we perform a series of experiments on two real-world datasets, with the aims of answering the following research questions:
\begin{itemize}
    \item \textbf{RQ1}: How does {\model} predict the interaction probability more precisely and converge to a better reranking strategy compared with SOTA models on the reranking task?
    \item \textbf{RQ2}: How do the well-designed components of {\model} (\eg Bi-LSTM, self-attention, etc.) influence the performance of {\model}?
    \item \textbf{RQ3}: How does the generator of {\model} provide a better reranking strategy intuitively?
    \item \textbf{RQ4}: How about the performance of the generator of {\model} in the real-world recommendation scenarios with efficient deployment?
\end{itemize}

\begin{table}
\caption{Statistics of datasets}
\label{table:dataset}
\begin{tabular}{lcc}
\toprule
{Description} & {Rec} & {Ad}\\
\midrule
{\#Users} & {2.16 $\times 10^8$} & {1.06 $\times 10^6$} \\
{\#Items} & {4.36 $\times 10^7$} & {8.27 $\times 10^5$} \\
{\#Records} & {3.01 $\times 10^9$} & {2.04 $\times 10^6$} \\
{\#User-item interactions} & {7.07 $\times 10^8$} & {2.04 $\times 10^7$} \\
{\#Avg size of $\mathcal{C}_u$} & {83.86} & {100.0} \\
{\#Avg size of $\mathcal{V}_u$} & {4.26} & {10.0} \\
\bottomrule
\end{tabular}
\end{table}

\subsection{Experimental Setup}
\subsubsection{Datasets}
We conduct extensive experiments on two real-world datasets: a proprietary dataset from Taobao application and a public dataset from Alimama, which are introduced as follows:
\begin{itemize}
    \item \textbf{Rec}~\footnote{https://www.taobao.com} dataset consists of large-scale list interaction logs collected from Taobao application, one of the most popular e-commerce platform in China. Besides, Rec dataset contains user profile (\eg id, age and gender), item profile (\eg id, category and brand), the input ranking list provided by the previous stages for each user and the labeled final ranking list. 
    \item \textbf{Ad}~\footnote{https://tianchi.aliyun.com/dataset/dataDetail?dataId=56} dataset records interactions between users and advertisements and contains user profile (\eg id, age and occupation), item profile (\eg id, campaign and brand). According to the timestamp of the user browsing the advertisement, we transform records of each user and slice them at each 10 items into list records. Moreover, we mix an additional 90 items with the final ranking list as the input ranking list to make it more suitable for reranking.
\end{itemize}
The detailed descriptions of the two datasets are shown in Table~\ref{table:dataset}. For the both datasets, we randomly split the entire records into training and test set, \ie we utilize 90\% records to predict the remaining 10\% interactions~\footnote{We hold out 10\% training data as the validation set for parameter tuning.}. 
 

\subsubsection{Baselines} 
We select two kinds of representative methods as baselines: point-wise and list-wise methods, which are widely adopted in most recommender systems. Point-wise methods (\ie DNN and DeepFM) mainly predict the interaction probability for the given user-item pair by utilizing raw features derived from user and item profile. List-wise methods (\ie MIDNN, DLCM , PRM and Seq2Slate) devote to extracting list-wise information with different well-designed principles.
The comparison methods are given below in detail:
\begin{itemize}
    \item \textbf{DNN}~\cite{Paul:YoutubeNet} is a standard deep learning method in the industrial recommender system, which applies MLP for complex feature interaction. 
    
    \item \textbf{DeepFM}~\cite{Guo:DeepFM} is a general deep model for recommendation, which combines a factorization machine component and a deep neural network component.
    
    \item \textbf{MIDNN}~\cite{Zhuang:MIDNN} extracts list-wise information of the input ranking list with complex handmade features engineering.
    
    \item \textbf{DLCM}~\cite{Ai:DLCM} firstly applies GRU to encode the input ranking list into a local vector, and then combine the global vector and each feature vector to learn a powerful scoring function for list-wise reranking.
    
    \item \textbf{PRM}~\cite{Pei:PRM} applies the self-attention mechanism to explicitly capture the mutual influence between items in the input ranking list.
    
    \item \textbf{Seq2Slate}~\cite{Bello:Seq2slate} applies rnn and pointer network to encode the previous selected items and select the most appropriate item at each step.
    
    \item \textbf{{\model}} is the novel context-wise framework proposed in this paper, which consists of the evaluator (\ie \textbf{\model}$_\mathcal{E}$) for predicting interaction probabilities and the generator (\ie \textbf{\model}$_\mathcal{G}$) for generating reranking results.
    
\end{itemize}
It is worthwhile to note that pair-wise and group-wise methods are not selected as baselines in our experiments due to their high training or inference complexities ( at least $\mathcal{O}(N^2)$) compared with point-wise ($\mathcal{O}(1)$) or list-wise and context-wise ($\mathcal{O}(N)$) models. Moreover, these methods usually achieve relatively worse performance, which have been reported in many previous studies~\citeN{Pei:PRM,Ai:DLCM,Burges:lambdamart}.

\subsubsection{Evaluation Metrics} 
To answer \textbf{RQ1}, we adopt different criteria to evaluate the model performance in following two aspects:
(1) To evaluate the accuracy of model predictions for {\model}$_\mathcal{E}$ and baselines, Loss (cross-entropy), AUC (area under ROC curve) and GAUC~\cite{Zhou:DIN} (average intra-user AUC) are applied.
(2) To evaluate the reranking results of {\model}$_\mathcal{G}$ and baselines, apart from NDCG (normalized discounted cumulative gain) metrics, we design the LR (list reward) metric to evaluate the overall context-wise profits of the whole list simulated by the evaluator (\ie {\model}$_\mathcal{E}$), which is calculated as follows: 
\begin{equation} 
    \begin{split}
    LR(\mathcal{O}) = &\sum_{x_o^i \in \mathcal{O}}\mbox{E}(x_o^i|u, \mathcal{O}).
    \end{split}
\end{equation}
Compared with NDCG, we argue that LR is a more fine-grained metric for the evaluation of rerank, which fully estimates the intra-correlations within the top-$k$ results and the overall profits are evaluated in a more precise manner. Moreover, the generalization capability of the LR metric is much stronger since it can rate the items that users have not browsed yet.

For online A/B testing in \textbf{RQ4}, We choose PV and IPV metrics, which are widely adopted in industrial recommender systems for evaluating online performance. Specifically, PV and IPV is defined as the total number of items that users browsed and clicked, respectively.


\subsubsection{Implementation}
We implement all models in Tensorflow 1.4. For fair comparison, pre-training, batch normalization and regularization are not adopted in our experiments. We employ random uniform to initialize model parameters and adopt Adam as optimizer using a learning rate of 0.001. Moreover, embedding size of each feature is set to 8 and the architecture of MLP is set to [128, 64, 32]. We run each model three times and reported the mean of results.

\subsubsection{Significance Test}
For experimental results in Tables 3, 4, 5 and 6, we use ``*'' to indicate that the best method is significantly different from the runner-up method based on paired t-tests at the significance level of 0.01.

\begin{table}
\caption{Overall performance comparison \wrt interaction probability prediction (bold: best; underline: runner-up).}
\label{table:model_performance_eva}
\setlength{\tabcolsep}{1.2mm}{
\begin{tabular}{lcccccc}
\toprule
\multirowcell{2}{Model} & \multicolumn{3}{{c}}{Rec} & \multicolumn{3}{{c}}{Ad} \\
\cmidrule(lr){2-4} \cmidrule(lr){5-7}
{} & {Loss} & {AUC} & {GAUC} & {Loss} & {AUC} & {GAUC} \\
\midrule 
{DNN} & {0.158} & {0.589} & {0.931} & {0.187} & {0.587} & {0.848} \\
{DeepFM} & {0.152} & {0.599} & {0.933} & {0.186} & {0.588} & {0.848} \\
\midrule
{MIDNN} & {0.143} & {0.610} & {0.936} & {0.185} & {0.600} & {0.848} \\
{DLCM} & {0.138} & {0.616} & {0.938} & {0.185} & {0.602} & {0.849}\\
{PRM} & {\underline{0.121}} & {\underline{0.630}} & {\underline{0.942}} & {\underline{0.184}} & {\underline{0.605}} & {\underline{0.850}} \\
{Seq2Slate$^\dag$} & {-} & {-} & {-} & {-} & {-} & {-}\\
\midrule
{{\model}$_{\mathcal{E}}$} & {\textbf{0.095}$^\ast$} & {\textbf{0.693}$^\ast$} & {\textbf{0.960}$^\ast$} & {\textbf{0.182}$^\ast$} & {\textbf{0.610}$^\ast$} & {\textbf{0.856}$^\ast$} \\
\bottomrule
\end{tabular}}
\begin{tablenotes}
\item[$\dag$] $^\dag$ Note that Seq2slate is designed to generate the final ranking list directly and can not be used to give predictions on the recorded final ranking list.
\end{tablenotes}
\end{table}

\begin{table}
\caption{Overall performance comparison \wrt the reranking strategy (bold: best; underline: runner-up).}
\label{table:model_performance_gen}
\begin{tabular}{l c ccc}
\toprule
\multirowcell{2}{Model} & \multicolumn{2}{{c}}{Rec} & \multicolumn{2}{{c}}{Ad} \\
\cmidrule(lr){2-3} \cmidrule(lr){4-5}
{} & {NDCG@5} & {LR@5} & {NDCG@5} & {LR@5} \\
\midrule 
{DNN} & {0.042}  & {0.164} & {0.109} & {0.199}  \\
{DeepFM} & {0.043}  & {0.169} & {0.111} & {0.203}  \\
\midrule
{MIDNN} & {0.050} & {0.175} & {0.117} & {0.212}  \\
{DLCM} & {0.052} & {0.182} & {0.119} & {0.228} \\
{PRM} & {\underline{0.054}} & {0.186} & {\underline{0.120}} & {0.232} \\
{Seq2Slate} & {0.050} & {\underline{0.192}} & {0.116} & {\underline{0.235}}  \\
\midrule
{{\model}$_{\mathcal{G}}$} & {\textbf{0.062}$^\ast$} & {\textbf{0.203}$^\ast$} & {\textbf{0.123}$^\ast$} & {\textbf{0.240}$^\ast$} \\
\bottomrule
\end{tabular}
\end{table}

\subsection{Performance Comparison (RQ1)}
We report the comparison results of the generator and evaluator module in {\model} (\ie {\model}$_\mathcal{G}$ and {\model}$_\mathcal{E}$) and other baselines on two datasets in Table~\ref{table:model_performance_eva} and Table~\ref{table:model_performance_gen}. The major findings from the experimental results are summarized as follows:

\begin{itemize}
    \item On both datasets, point-wise methods (\ie DNN and DeepFM) achieve relatively pool performance for the task of  interaction probability prediction and reranking.
    It indicates that feature engineering of user and item profile is incapable of covering the list-wise information contained by the input ranking list. Generally, list-wise methods achieves remarkable improvements in most cases. By considering the mutual influence among the input ranking list, these methods learn a refined scoring function aware of the feature distributions of the input ranking list. Among these methods, DLCM and PRM consistently outperform MIDNN in all cases, proving that deep structures (\ie RNN and self-attention) has superior abilities in extracting the feature distribution  compared with handmade feature engineering. Moreover, compared with DLCM, adequate improvements are observed by PRM, which demonstrates the effectiveness of self-attention and personalization in the reranking task.

    
    \item {\model}$_{\mathcal{E}}$ consistently yields the best performance on the Loss, AUC and GAUC metrics of both datasets. 
    In particular, {\model}$_{\mathcal{E}}$ improves over the best baseline PRM by 0.073, 0.022 in AUC, and 0.005, 0.006 in GAUC on Rec and Ad dataset, respectively. By taking full advantage of the contextual information in the final ranking list, {\model}$_{\mathcal{E}}$ shows the excellent ability to provide more precise contextual interaction probabilities. Different from extracting information from the disordered input ranking list in DLCM and PRM, with the help of well-designed Bi-LSTM and self-attention mechanism, {\model}$_{\mathcal{E}}$ capture the sequential dependency and mutual influence in the final ranking list, which is another essential and effective factor to affect the contextual predictions.
    
    
    
    
    \item {\model}$_{\mathcal{G}}$ significantly and consistently outperforms state-of-the-arts methods by a relatively large margin on both datasets across all metrics. Overall, on Rec and Ad dataset,  {\model}$_{\mathcal{G}}$ achieves performance gains over the best baseline (\ie Seq2Slate) by 0.008 and 0.003 for NDCG , 0.011 and 0.005 for LR, respectively, which evidences that the technically designed model architecture and training procedure of  {\model}$_{\mathcal{G}}$ have successfully transformed the context-wise ability of {\model}$_{\mathcal{E}}$ into production. The evolving and activating layers model the information of the selected list, assisting the selector layer to make the most appropriate choice for each step by comparing in the input ranking list. Besides, under the guidance of {\model}$_{\mathcal{E}}$, policy gradient with the proposed advantage reward helps distill the reranking knowledge into {\model}$_{\mathcal{G}}$ comprehensively. 
    
    
    
    \item Compared with the experimental results on the Rec dataset, the performance lift on the Ad dataset is relatively slight. One possible reason is that Ad dataset is published with random sampling, resulting in the inconsistent and incomplete list records as well as  weaker intra-list relevance of user feedback.
\end{itemize}

\subsection{Study of {\model} (RQ2)}

\begin{table}
\caption{Ablation study of the evaluator (BL: Bi-LSTM; SA: self-attention).}
\label{table:ablation_study_eva}
\begin{tabular}{l c cc}
\toprule
{Model} & {Loss} & {AUC} & {GAUC} \\
\midrule
{{\model}$_{\mathcal{E}}$( - BL )} & {0.120} & {0.632} & {0.942} \\
{{\model}$_{\mathcal{E}}$( - SA )} & {0.102} & {0.684} & {0.955}\\
\midrule
{{\model}$_{\mathcal{E}}$} & \textbf{0.095$^\ast$} & \textbf{0.693$^\ast$} & \textbf{0.960$^\ast$} \\
\bottomrule
\end{tabular}
\end{table}

\begin{table}
\caption{Ablation study of the generator (EL: evolving layer; AL: Activating layer; DR: different reward; SR: self reward).}
\label{table:ablation_study_gen}
\begin{tabular}{lcc}
\toprule
{Model} & {NDCG@5} & {LR@5} \\
\midrule 
{Greedy} & {0.054} & {0.184}\\
{ {\model}$_{\mathcal{G}}$( - EL)} & {0.060} & {0.192}\\
{ {\model}$_{\mathcal{G}}$( - AL) } & {0.055} & {0.185}\\
{ {\model}$_{\mathcal{G}}$( - DR) } & {0.059} & {0.195}\\
{ {\model}$_{\mathcal{G}}$( - SR) } & {0.033} & {0.099}\\
\midrule
{{\model}$_{\mathcal{G}}$} & {\textbf{0.062$^\ast$}} & \textbf{0.203$^\ast$} \\
\bottomrule
\end{tabular}
\end{table}

In this section, we perform a series of experiments on the Rec dataset to better understand the traits of {\model}, including well-designed components of evaluator, generator and training procedure. It is noteworthy that similar trends can also be observed on the Ad dataset, which are omitted due to the page limitation.

\subsubsection{Ablation Study of {\model}$_{\mathcal{E}}$} By leveraging the contextual information in the final ranking list, {\model}$_{\mathcal{E}}$ is designed to predict the context-wise interaction probability more precisely. To examine the effectiveness of in the interacting layer, we prepare two variants of {\model}$_{\mathcal{E}}$:
\begin{itemize}
    \item {\model}$_{\mathcal{E}}$( - BL): The variant of evaluator, which removes the Bi-LSTM.
    \item {\model}$_{\mathcal{E}}$( - SA): The variant of evaluator, which removes the self-attention mechanism.
\end{itemize}

The comparison results of {\model}$_{\mathcal{E}}$ with its variants are show in Table \ref{table:ablation_study_eva}. The declines of AUC and GAUC metrics are observed after removing each component, which demonstrate their effectiveness for capturing the context-wise information. Besides, {\model}$_{\mathcal{E}}$( - BL) performs worse than {\model}$_{\mathcal{E}}$( - SA), which means modeling the two-way sequential information of the final ranking list is more important.

\subsubsection{Ablation Study of Generator and Training Procedure} 
Both the model design and training procedure help achieve the best reranking results. In this section, we carefully review the contributions of each component. The variants of generator and training procedure are listed as follows:
\begin{itemize}
    \item Greedy: Directly reranking in descending order according to the score of evaluator. 
    \item {\model}$_{\mathcal{G}}$( - EL): The variant of generator without evolving layer, which ignores  the evolution of user intent (\ie Eq.~\ref{eq:gen-gru}).
    \item {\model}$_{\mathcal{G}}$( - AL): The variant of generator without activating layer, which ignores the influence between items in the input ranking list (\ie Eq.~\ref{eq:gen-att}).
    \item {\model}$_{\mathcal{G}}$( - DR): The variant of training procedure, which removes the differential reward (\ie Eq.~\ref{eq:advantage-reward-2}) in the advantage reward.
    \item {\model}$_{\mathcal{G}}$( - SR): The variant of training procedure, which removes the self reward (\ie Eq.~\ref{eq:advantage-reward-1}) in the advantage reward.
\end{itemize}

We summarize the results in Table~\ref{table:ablation_study_gen} and have the following observations: 
\begin{itemize}
    \item Reranking based the greedy strategy achieves a relatively poor performance on the on both  metrics.
    It indicates that the {\model}$_{\mathcal{E}}$ is capable of capturing context-wise information, while it may not be a good way to ranking directly by scores since the greedy strategy changes each item's context in the final ranking list.
    \item The performance of {\model}$_{\mathcal{G}}$( - EL) and {\model}$_{\mathcal{G}}$( - AL) proves the importance of modeling the selected item list for a better choice at each step.
    \item Intuitively, {\model}$_{\mathcal{G}}$( - SR) perform worst due to the lack of item's own reward in the list. {\model}$_{\mathcal{G}}$ performs better than {\model}$_{\mathcal{G}}$( - DR) by considering the context-wise reward of each item in the final ranking list more comprehensively. 
\end{itemize}

\begin{figure}
    \centering
    \includegraphics[scale=0.22]{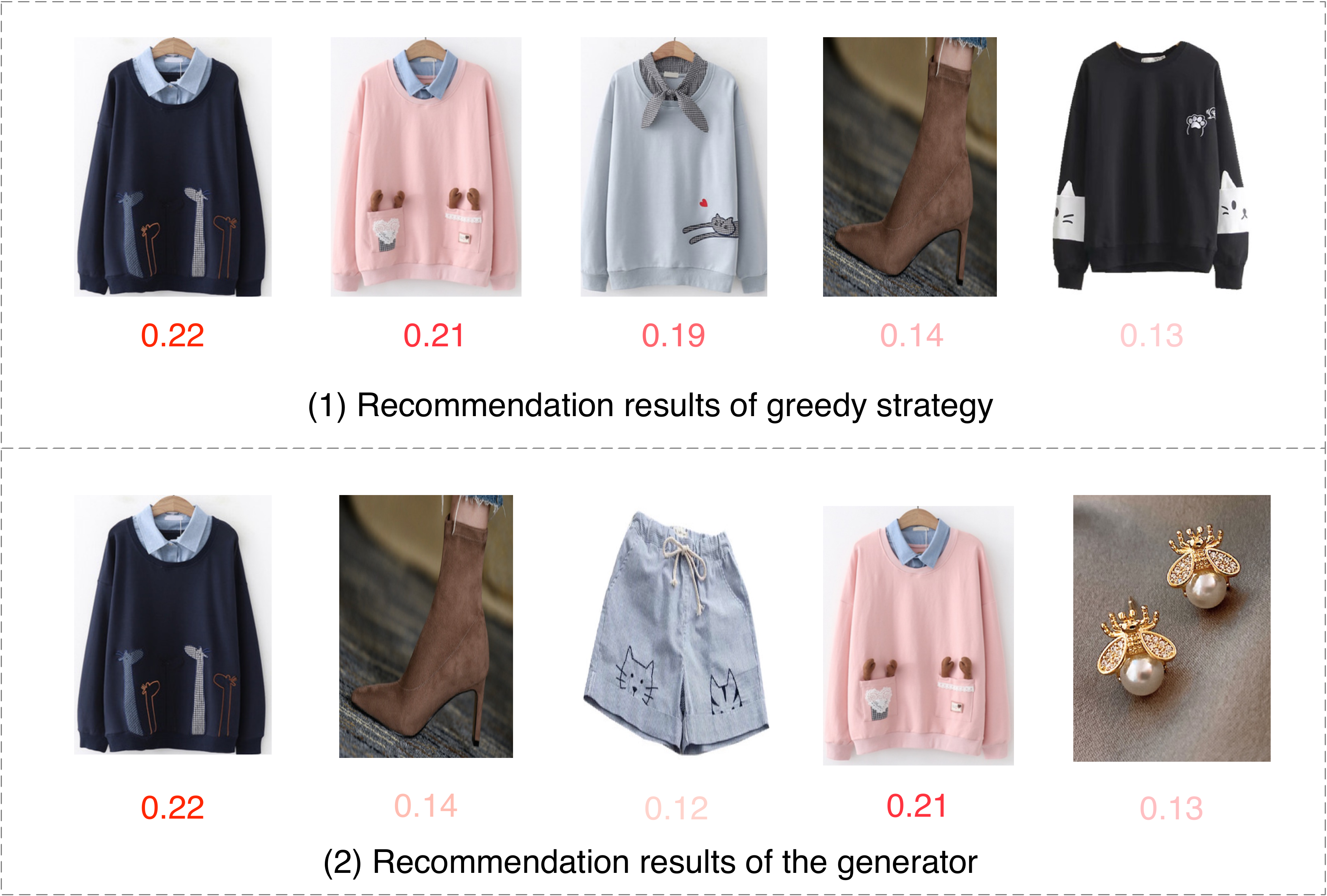}
    \caption{Illustrative example from Rec dataset, which shows the difference between greedy strategy and our learned reranking strategy.}
    \label{fig:case_study}
\end{figure}

\subsection{Case Study (RQ3)}
In order to clearly demonstrate  how {\model} addresses the limitations of the greedy reranking strategy existed in previous methods, we conduct one case study in the large-scale industrial Rec dataset. As shown in Fig.~\ref{fig:case_study}, the main findings are summarized as follows:
\begin{itemize}
    \item The rating scores below the item are estimated by a well-trained point-wise model. Intuitively, according to the score distribution, we can find that the target user is most interested in clothing category recently, followed by jewelry and pants.
    
    \item The greedy reranking strategy choose to place four clothes in its top-5 results. After this user thoroughly browsing and engaging with the first clothe, he/she may be tired of the clothing category and eager for other categories he/she is also interested in. In this case, the greedy reranking strategy can not reach the best recommendation results and further decrease the user experience.
    
    \item Different form the greedy reranking strategy, the generator selects the item by considering the contextual items. The recommendation results are more pluralistic, taking more chance to capture and activate the user's latent interests. Hence, the generator shows its ability to generate a more contextually effective and attractive recommendation list, making users engage with the RS more.
\end{itemize}

\begin{figure}
    \centering
    \includegraphics[width = 9cm]{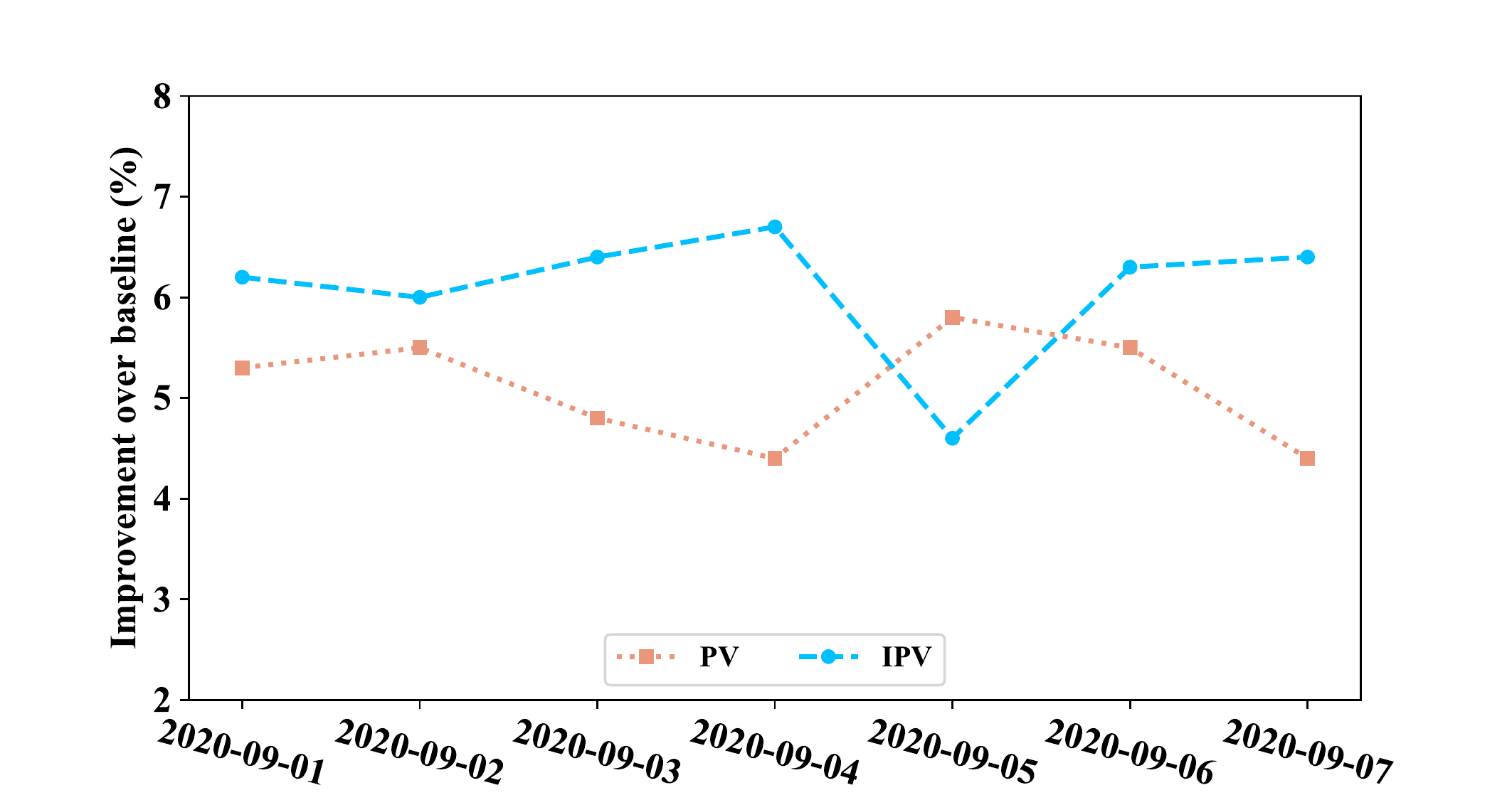}
    \caption{Online performance on the homepage for the \emph{Guess You Like} scenario in Taobao app. $y$-xis denotes the improvement ratio over the existed deployed baseline.}
    \label{fig:online_performance}
\end{figure}

\subsection{Online A/B Testing (RQ4)}
To verify the effectiveness of our proposed framework {\model} in the real-world settings, {\model} has been fully deployed in homepage for the  \emph{Guess You Like scenario}, the most popular recommendation scenario of Taobao application~\footnote{In practice, we only deploy the {\model}$_\mathcal{G}$ into production for serving reranking task.}. In this waterfall scenario, users can browse and interact with items sequentially and endlessly. The fixed-length recommended results are displayed to the user when he/she reaches the bottom of the current page. Considering the potential issue of high latency to the user experience, it brings great challenge to deploy {\model} into production. 

To this end, we make the following improvements for efficient deployment: 
\begin{itemize}
    \item Instead of the traditional cloud-to-edge framework, we directly deploy {\model} on edge as introduced in EdgeRec~\cite{Gong:EdgeRec}. In this way, the network latency between the cloud server and user edge is saved. Nearly 30 milliseconds are saved, consisting of 10 requests and 3 milliseconds per request, which could be a bottleneck of deploying a context-wise algorithm.
    
    \item Considering the long total time cost, we developed the advance trigger and delay completion mechanism. Specifically, {\model} is requested a few places in advance of the last position of current page. Besides, {\model} will return more results than expected at each request, to make up for the vacancy of the next request.
\end{itemize}
After successful deployment of the proposed {\model}, we evaluate the performance from ``2020-09-01'' to ``2020-09-07'' and report the performance comparison with deployed baseline model PRM in the Fig.~\ref{fig:online_performance}. Not surprisingly, we observe that {\model} consistently and significantly outperforms PRM model across both PV and IPV metrics, which further verifies the effectiveness of our proposed framework {\model}. Overall, {\model} achieves performance improvement over the best baseline PRM of \textbf{5.2}\% for PV and \textbf{6.1}\% for IPV, with the average cost of \textbf{32} milliseconds for online inference.

\section{Conclusion}
In this paper, we highlight the importance of modeling the contextual information in the final ranking list and address how to leverage and transform such information for better recommendation. We propose a novel two-stage context-wise framework {\model} to learn to rank with contexts. Furthermore, we elaborate on the model design of evaluator and generator in {\model}, which aim to predict the contextual interaction probability more precisely and learn a better context-wise reranking strategy, respectively. Extensive experiments on both industrial and benchmark datasets demonstrate the effectiveness of our framework compared to state-of-the-art point-wise and list-wise methods. Moreover, {\model} has also achieved impressive improvements on the PV and IPV metrics in online experiments after successful deployment in one popular recommendation scenario of Taobao application.
In the future, we will investigate into how to incorporate the long-term reward for learning a  better reranking strategy.



\bibliographystyle{ACM-Reference-Format}
\balance
\bibliography{references}

\end{document}